\documentstyle[11pt,gh2001-asp,twoside,psfig]{article}
\def\edcomment#1{\iffalse\marginpar{\raggedright\sl#1\/}\else\relax\fi}
\reversemarginpar

\begin{document}
\title{Scattering of Stars by Transient Spiral Waves}
 \author{J. A. Sellwood and Miguel Preto}
\affil{Rutgers University, Department of Physics \& Astronomy, \\
136 Frelinghuysen Road, Piscataway, NJ 08854-8019, USA}

\begin{abstract}

Transient spiral waves of moderate amplitude cause substantial changes to 
the angular momenta of many stars in a galaxy disk.  Stars near to corotation 
are affected most strongly: for a wave of $\sim 20\%$ overdensity, the rms change for particles near to corotation is $\sim18\%$ of their initial angular 
momenta.  Yet these changes produce no increase in random 
motion near corotation, while the same wave causes mild heating near its Lindblad resonances.  The direction of 
radial migration depends both on initial location relative to corotation and on azimuth.  Streamlines of cold gas become very significantly distorted before they cross, suggesting that gas is also mixed radially by this mechanism.  Such continuous churning of the disk has profound implications for 
abundance gradients in galaxies.  Furthermore, it is quite possible that the Sun has 
migrated significantly from its radius of birth, consistent with the claims of Wielen {\it et al}.
\end{abstract}

\section{Introduction}
Barbanis \& Woltjer (1967) first showed that a spiral wave disturbance of 
finite time duration will, on average, scatter stars away from circular orbits. 
 Many subsequent $N$-body simulations ({\it e.g.}, Sellwood \& Carlberg 1984) 
have confirmed that transient spiral waves do heat the disk.  Earlier, 
Spitzer \& Schwarzschild (1953) had shown that stars are also scattered by 
dense objects, such as giant molecular clouds, co-orbiting within the disk.

The theory of these scattering processes has been developed in many 
subsequent papers, and it is now generally agreed that: (1) transient spirals 
heat mostly the in-plane motions, while (2) giant molecular clouds do little 
heating themselves but act mostly as scatterers which try to isotropize 
peculiar velocities by converting some of the in-plane components to 
vertical.

Scattering by spiral waves is still not completely understood.  
The 
cumbersome epicyclic approach by Barbanis \& Woltjer was supplanted by 
the 
more powerful action-angle machinery introduced by Lynden-Bell \& Kalnajs 
(1972).  Subsequent work by Dekker (1976), Carlberg \& Sellwood (1985), and 
Fuchs (2001) has developed a formalism which is somewhat related to 
quasi-linear theory in plasma physics.  Binney \& Lacey (1988) derived 
diffusion coefficients in action space while Jenkins \& Binney (1990) used 
solar 
neighbourhood data to determine the relative efficiency of scattering 
by 
clouds and by spirals.
  Little \& Carlberg (1991) report a numerical study of the angular momentum changes of test particles under the influence
 of a transient, weak, imposed bar and report results in general agreement with the predictions of Lynden-Bell \& Kalnajs.

\begin{figure}[t]
\centerline{\psfig{figure=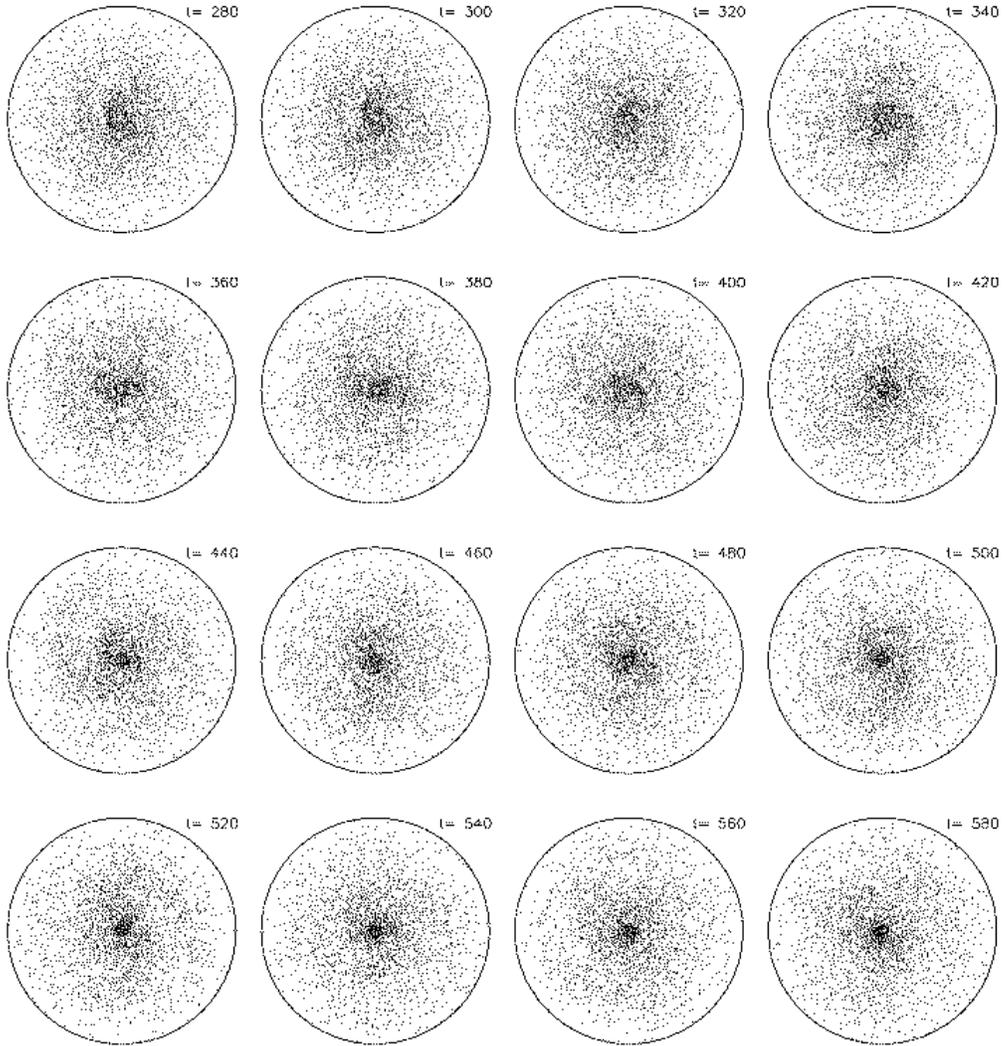,width=\hsize,angle=0,clip=}}
\caption{Part of the evolution of a simulation which supports many very mild, 
transient spiral features.  The radius of the circles is $21R_0$ and only one 
particle in 20 is plotted.}
\end{figure}

The conclusions from all this work correctly predict {\it orbit averaged\/} 
changes caused by spirals: small net changes are expected at corotation, 
while larger non-zero net changes occur at Lindblad resonances.  But it is 
not generally appreciated that typical changes at corotation due to a 
moderate amplitude transient spiral are {\it large}, and spirals can cause 
stars to migrate over substantial radial 
distances -- in fact, several papers 
have concluded the opposite!  Wielen ({\it 
e.g.} Wielen {\it et al.}\ 1996) 
is a notable exception to this 
generalization; he has long suspected that the 
Sun, because of its atypically high metallicity, has migrated some 2kpc outwards from its 
place of birth.

Here we report our on-going studies of scattering by spiral waves in $N$-body simulations, and describe the range of possible changes for individual 
particles.  In particular, we find that transient spirals of moderate 
amplitude cause surprisingly large angular momentum changes, but no significant heating, in the vicinity 
of corotation.

\begin{figure}[t]
\centerline{\psfig{figure=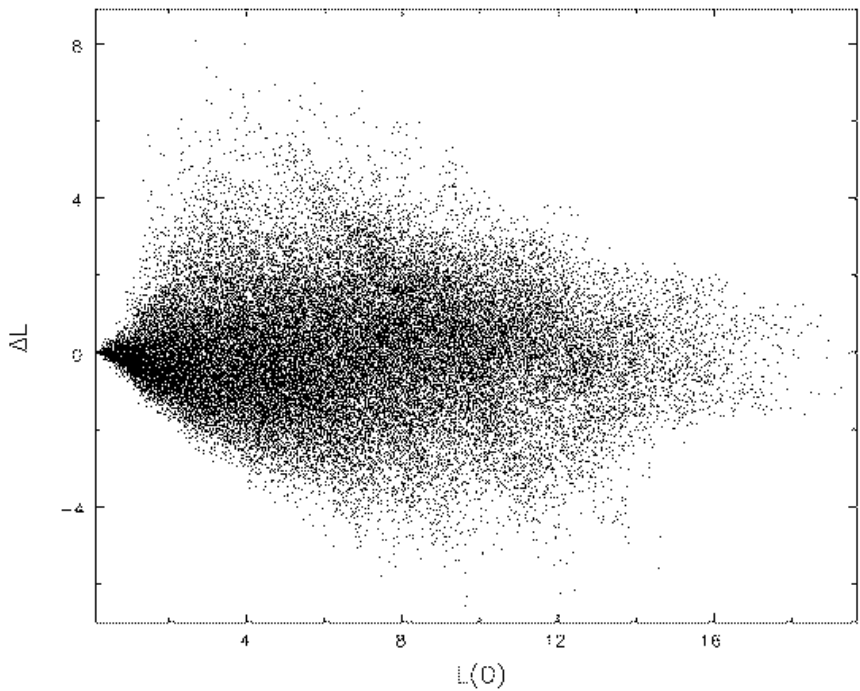,width=0.9\hsize,angle=0,clip=}}
\caption{The changes in $L_z$ of individual particles between $t=0$ and 
$t=800$ in the simulation shown in Figure 1.  Only one particle in 19 is 
plotted.}
\end{figure}

\section{Simulation}

Figure 1 shows the evolution of a half-mass, $V_0=\,$const., or Mestel, disk which is gently tapered at its inner and outer edges.  We set the parameter of the distribution function ({\it e.g.}\ Binney \& Tremaine 1987; eq 4-162) to be $q=11.44$ in order that $Q=1.5$ in the untapered disk. 
 Toomre (1981) predicts this 
disk is stable if smooth, but particle noise in 
simulations (which is also present in real galaxy disks!)\ always induces 
some spiral activity.  The $N=1\,$M particles of the simulation are confined 
to a plane and their mutual gravitational forces are determined with the aid of a 2-D polar 
grid.  Forces from azimuthal wavenumbers $0 \leq m \leq 4$ only are included, and are derived from the usual Plummer softening law, with scale length $0.1R_0$.  We use 
$V_0$, the constant circular speed of the Mestel disk, as our velocity unit, 
and the mean radius of the inner taper $R_0$ as our length unit; thus our units of 
time and angular momentum are $R_0/V_0$ and $R_0V_0$ respectively.

Figure 2 shows the change in angular momenta from $t=0$ to $t=800$ of a 
representative sample of the particles.  The changes are generally quite 
large (rms $\sim 2.68$), even though the spiral activity was very mild; non-axisymmetric variations in surface density exceed 10\% for only two or three patterns.  We will 
show later that the angled features are signatures of scattering at 
corotation.  Closer analysis indicates that most particles are affected by 
more than one wave over any short period.  Evidently, the changes shown in 
Figure~2 are far too complicated to understand in detail.

\begin{figure}
\centerline{\psfig{figure=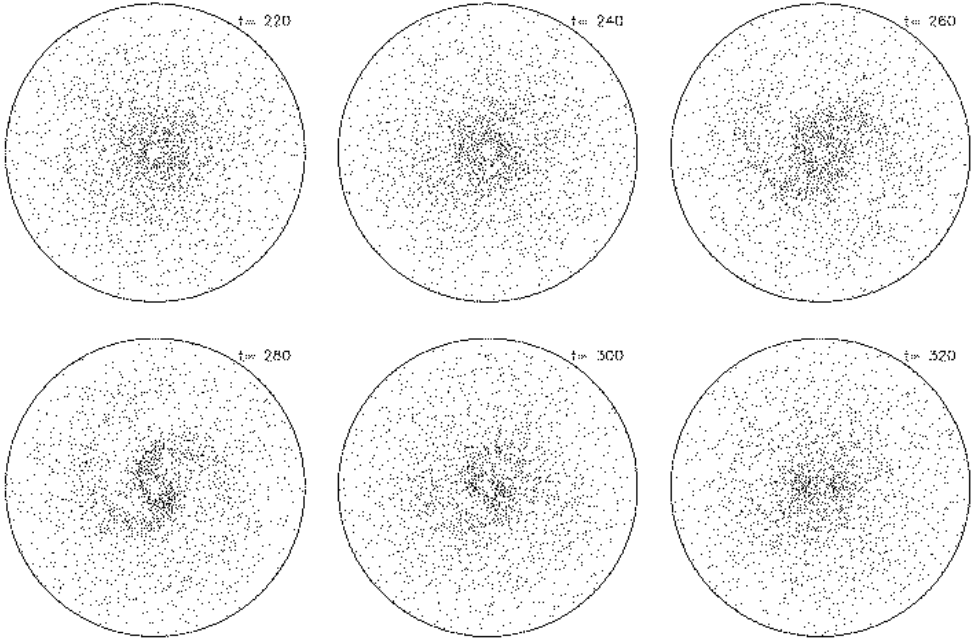,width=\hsize,angle=0,clip=}}
\caption{The later part of the evolution of a simulation at supports an 
isolated spiral.  The spiral has largely decayed by $t=300$, but a weak bar 
is then visible in the inner disk.  The radius of the circles is $15R_0$ and the groove that seeded the spiral wave is centered at $L_z=6.5$.}
\end{figure}

\section{An ``isolated'' spiral}
In order to have a chance to understand this behaviour, we turn to a much 
more restricted simulation which was deliberately crafted to excite an 
isolated spiral mode that stands out clearly from the noise.  It uses the 
same $Q=1.5$, half-mass Mestel disk, but particle noise was suppressed by 
using a quiet start (Sellwood 1983), disturbance forces were restricted to 
$m=2$ only, and the single spiral pattern was seeded by a groove (Sellwood \& 
Kahn 1991).
  The groove in angular momentum density is centered on $L_z=6.5$ and has a half-width in $L_z$ of 0.2.

The later part of the evolution of this simulation is shown in Figure 3.  The 
single transient spiral pattern stands out well, although the evolution is, 
unfortunately, tainted at the end by a weak bar pattern that appears as the 
spiral decays.  The bar feature seems unavoidable.
  Note that since disturbance forces include only the $m=2$ component, this simulation does not capture the full non-linear evolution.

\begin{figure}[t]
\centerline{\psfig{figure=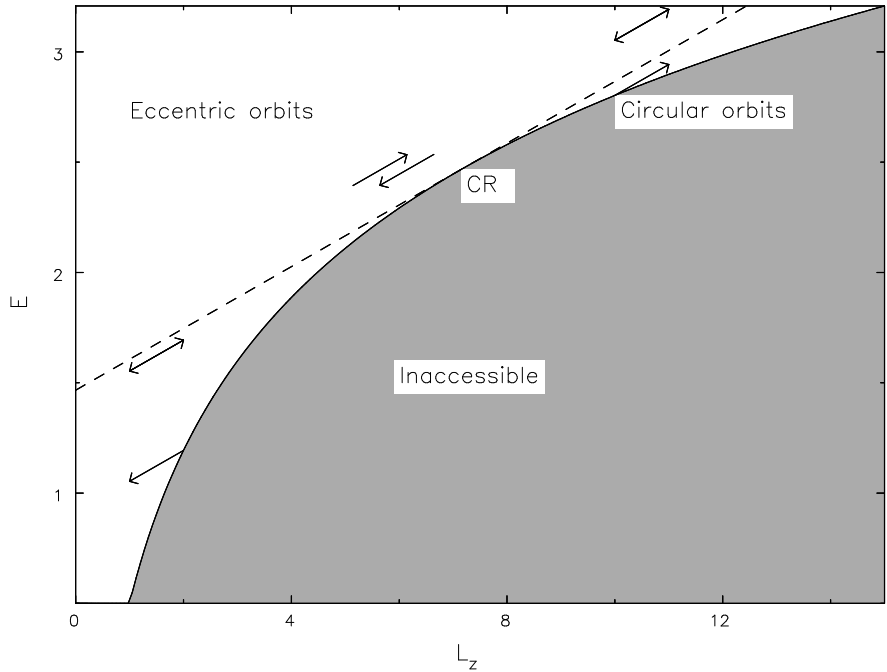,width=0.9\hsize,angle=0,clip=}}
\caption{The Lindblad diagram.  The circular velocity curve separates the 
inaccessible shaded area from the region populated by eccentric orbits.  The 
arrows indicate a number of possible scatterings of stars by a steadily 
rotating wave which has a pattern speed equal to the slope of the tangent at 
corotation (dashed).}
\end{figure}

\subsection{Standard results for a steady wave}
Jacobi's invariant, $E_J = E - \Omega_p L_z$, is a conserved quantity 
for a test particle in a steadily rotating non-axisymmetric potential.  
Therefore, changes in 
angular momentum and energy are related as
\begin{equation}
\Delta E = \Omega_p \Delta L_z.
\end{equation}

It is helpful to draw the classical Lindblad diagram to understand scattering 
of particles by a steadily-rotating, non-axisymmetric perturbation.  The 
solid curve in Figure~4 is the curve for circular orbits which separates the 
inaccessible region of the diagram (shaded) from the region where orbits are 
eccentric.  The energy of random motion is the vertical distance from the 
circular orbit curve.

Because of equation (1), particles are scattered along 
lines of constant slope, $\Omega_p$.  In general, scattering along a line of 
slope $\Omega_p$ leads to changes in the relative amounts of orbital to 
epicyclic energy for a particle, but 
because the slope is also the slope of 
tangent to the circular orbit curve at corotation, changes at corotation do 
not change the random energy of a particle to first order.   Finally, there 
is a bias in the possible shifts, which must be away from circular orbits 
when not at corotation.
  This bias away from circular orbits is the root 
cause of heating by spiral waves away from corotation.

\begin{figure}
[t]
\centerline{\psfig{figure=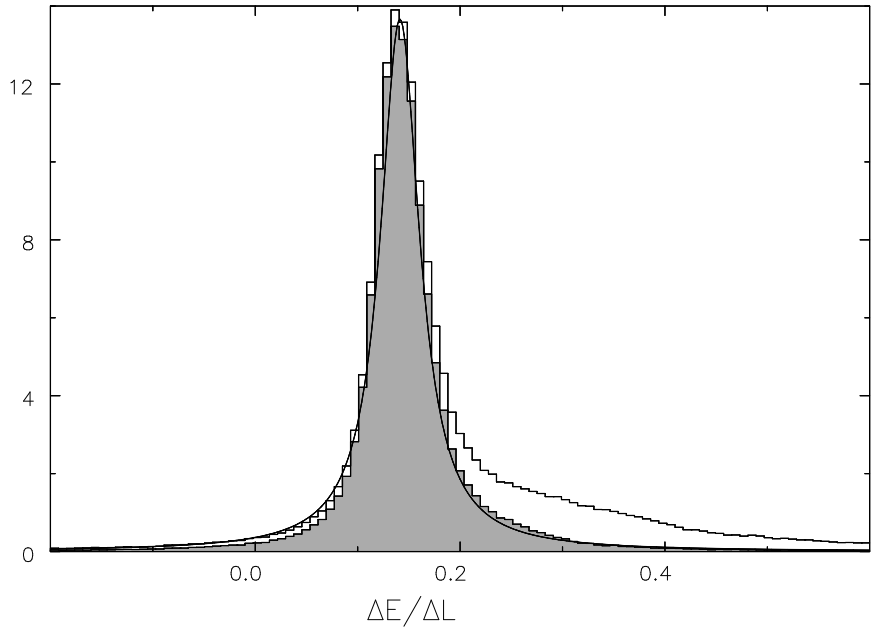,width=0.9\hsize,angle=0,clip=}}
\caption{Histogram of the ratio $\Delta E/\Delta L_z$ measured from the particles over the time interval to $t=300$.   The shaded histogram is for particles which start with $L_z>2$ while the unshaded includes all particles with significant $\Delta L_z$.  The curve is 
equation (3) with the pattern speed and growth rate measured from the simulation.}
\end{figure}

\subsection{Effect of finite growth}
The spiral wave in the simulation grows exponentially, saturates and decays, implying that the perturbing potential does not have a single pure frequency as assumed in equation (1).  To see the effect of the finite duration of a growing wave, we approximate the perturbing potential as a purely growing mode
\begin{equation}
\Phi_1(r,\phi) e^{im\Omega_p t}e^{\gamma t} \times H(t)[1-H(t-T)],
\end{equation}
where the Heaviside functions limit the time interval, $0<t<T$, over which the perturbation is non-zero.  Here, $\Omega_p$ is the pattern speed of the $m$-armed mode, $\gamma$ is its growth rate, and $m=2$ in our case.  The power spectrum of this perturbation is an expression with three terms, but once $\gamma T \gg 1$, the dominant term has a simple Lorentzian form; in this limit, the power spectrum, normalized to integrate to unity, is
\begin{equation}
\left|\Phi_\Omega(r,\phi)\right|^2 \simeq {m \over \pi}\; {\gamma \over m^2(\Omega - \Omega_p)^2 + \gamma^2}
.
\end{equation}

Particles in the the simulation experience a spread of frequencies given by this expression, and thus the ratio $\Delta E/\Delta L_z$ measured from the particles should also have this spread.  Accordingly, we determined $\Delta E$ and $\Delta L_z$ separately for every 
particle in the simulation between the start and $t=300$ and histogram their 
ratio in Figure~5; we have discarded particles with $\Delta L_z/L_z<0.01$ for which $\Delta E/\Delta L_z$ is less precisely determined.  The shaded histogram is for particles with initial 
$L_z>2$ while the unshaded is for all particles.
  Note that particles with $L_z<2$ are all inside the inner Lindblad resonance of the 
spiral.

The smooth curve in Figure~5 is equation (3) with the parameters $\Omega_p = 0.141$ and $\gamma 
= 0.047$ measured from the simulation -- {\it 
i.e.}\ the curve is not fitted 
to the histogram, but the curve and the shaded histogram are 
normalized to enclose 
the same integrated area.
  The correspondence between the prediction and the empirical values clearly shows that 
the large majority of particles with $L_z>2$ is affected only by the one 
pattern.  The curve is not quite a perfect fit to the shaded histogram; some discrepancies are due to the fact that the wave saturated and decayed before $t=300$.

The scattering trajectories shown in Figure 4 assume a steadily rotating, constant amplitude pattern, but because the frequency is broadened by the finite growth time of the mode, particles will be 
scattered along lines having the range of slopes spread according to equation (3) about the mean, which has slope $\Omega_p$.

\begin{figure}[t]
\centerline{\psfig{figure=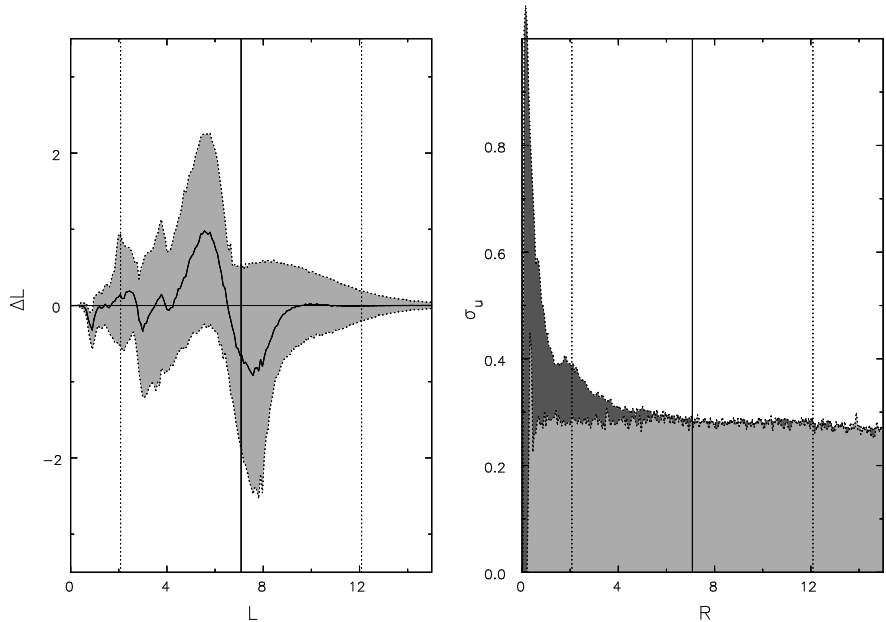,width=.9\hsize,angle=270,clip=}}
\caption{Left: The mean (solid curve) and $\pm 1\sigma$ (shaded) variation of 
$L_z$ between $t=0$ and $t=300$ for the simulation shown in Figure 3.  The 
vertical lines mark corotation (solid) and the Lindblad resonances (dotted). 
Right: The light shaded area indicates the rms radial speed ($\sigma_u$) of 
particles at $t=0$ and the dark shaded region is the increase in this 
quantity by $t=300$.  Notice that the substantial changes in $L_z$ produce almost no heating n the vicinity of corotation.}
\end{figure}

\begin{figure}[t]
\centerline{\psfig{figure=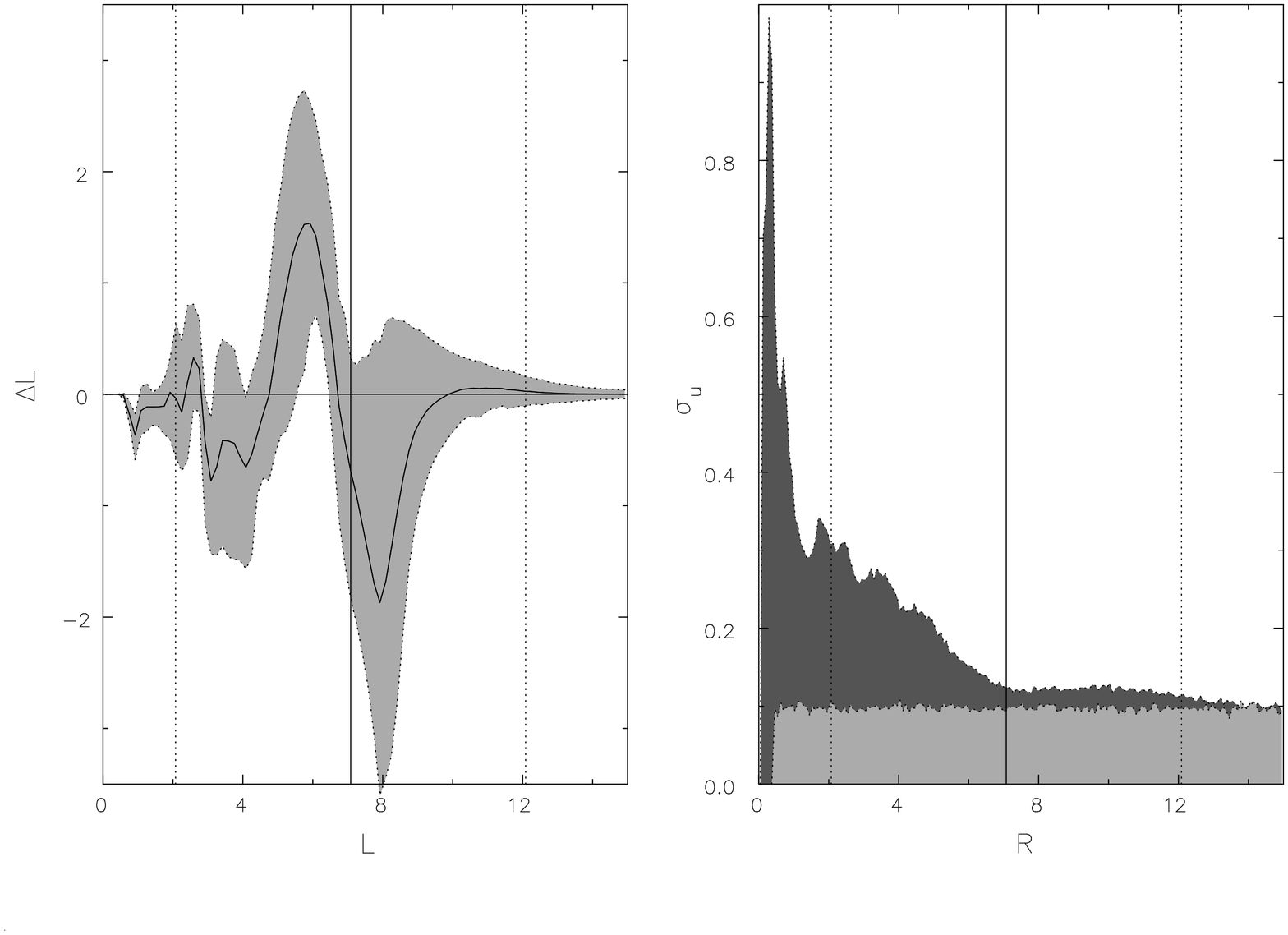,width=.9\hsize,angle=270,clip=}}
\caption{The same as for Figure 6, except that only 20\% of the particles 
with the smallest epicycle energies contribute to these plots.}
\end{figure}

\subsection{Scattering by one wave}
Figure~6 (left) shows the mean (line) and $\pm 1\sigma$ distribution of 
$\Delta L_z$ as a function of initial $L_z$.  Note that these changes are 
azimuthally averaged.  Corotation and the two Lindblad resonances are marked. 
 Most changes occur near corotation, but with huge scatter; most particles 
inside corotation gain $L_z$ while most outside lose $L_z$.   There are 
roughly as many gainers as losers, while the average change for each particle 
is large enough that most cross this resonance; in effect, particles 
generally exchange places across corotation.  Figure~6 (right) confirms that 
the large changes at corotation cause very little heating; 
significant heating occurs only in the inner disk, where it is very difficult to disentangle the contributions from 
the inner Lindblad resonance (ILR) and from the bar.

Figure~7 shows the same plots but now for the $\sim 20\%$ of particles that 
start with the smallest epicycle energies.  For these, the changes around 
corotation are even larger, but now one can see a clear excess of losers 
between ILR and corotation and a small excess of gainers between corotation and outer Lindblad resonance (OLR).  The 
panel on the right shows the increase in rms radial velocity of the same 
particles; some Lindblad resonance heating is noticeable, although this subset of the particles may not reveal the full story.
  Changes in both panels of Figure~7 are larger inside corotation than outside because the wave amplitude is appreciably greater inside than outside.

Note that the greatest changes occur on opposite sides of corotation for particles several times farther from the groove center than the groove half-width ($0.2$ in $L_z$).  Thus, these changes are produced by the spiral response, and are not a feature caused by the narrow groove we used to seed the instability.

\begin{figure}
\centerline{\psfig{figure=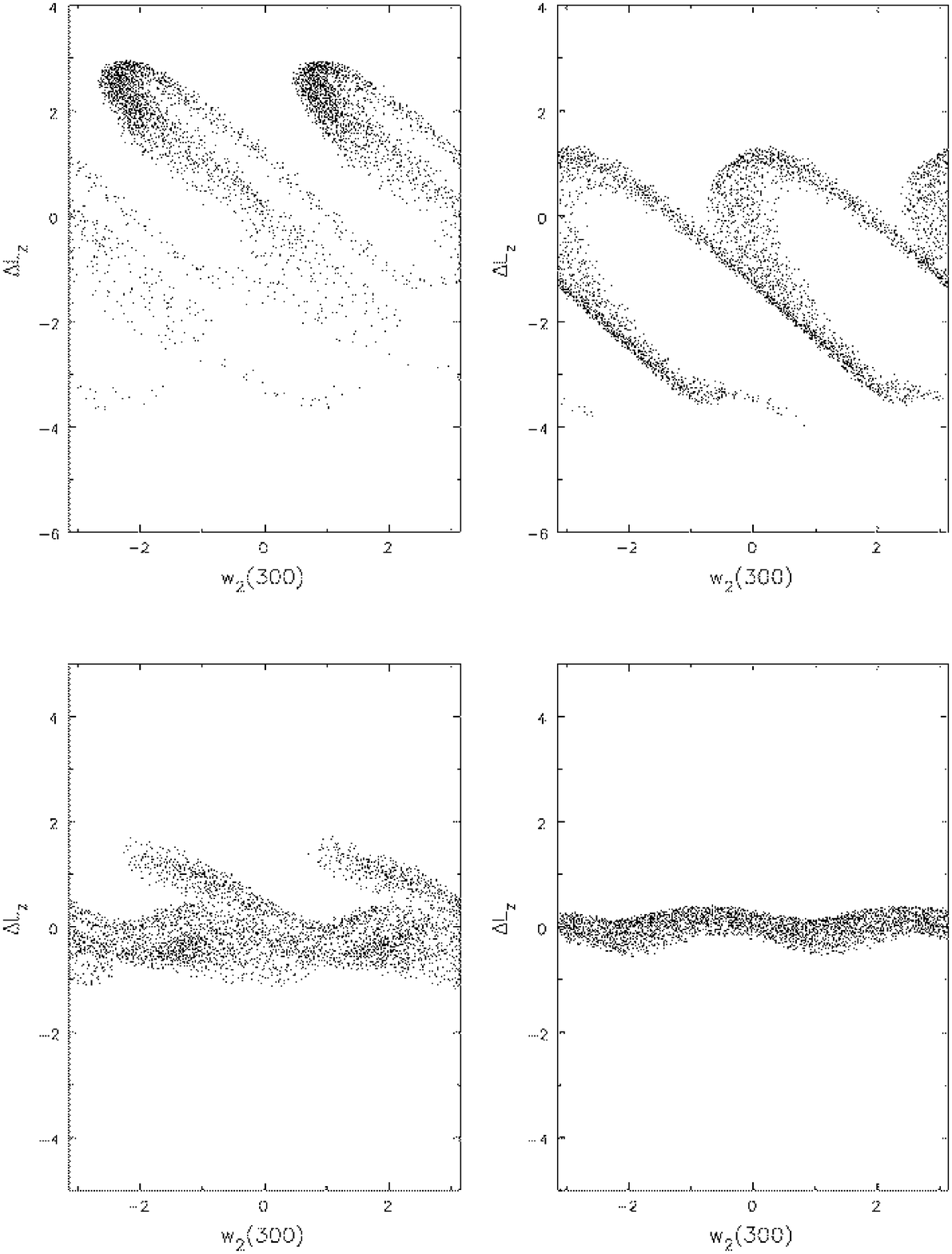,width=\hsize,angle=0,clip=}}
\caption{The change in $L_z$ as a function of azimuthal angle $w_2$ for 
particles starting with initial $L_z$ in four different narrow ranges.  In 
each panel, the range of initial $L_z=0.3$ centered on: top left 5.85, top 
right 7.65, bottom left 2.0, bottom right 12.0.  Some 56\% of particles with 
the largest epicycle energies at $t=0$ have been omitted.}
\end{figure}

\subsection{Azimuthal variations}
Figure~8 shows $\Delta L_z$ for individual particles as a function of their 
azimuthal angle at $t=300$.  The angle, $w_2$, is the variable conjugate to the angular momentum (Lynden-Bell \& Kalnajs 1972), and is 
determined from the instantaneous 
position of the particle assuming an undisturbed, axisymmetric potential.  
The four panels show different groups of particles whose initial angular 
momenta placed them just inside or outside corotation (top), near the ILR and 
the OLR (bottom).  Only the $\sim44$\% of particles with smallest epicyclic energies are 
plotted; the patterns in these panels are less marked when all particles are plotted.  The obvious double period is to be 
expected for a two-arm pattern.

The coherence of the changes with phase relative to the wave is striking.  Particles are being moved in a systematic manner by the large-scale potential perturbation, to produce these striking ribbons in phase space.  It is clear that a large part of the spread 
at corotation seen in Figure~6 is not random scatter, but is strongly 
correlated with azimuth.

\begin{figure}
[t]
\centerline{\psfig{figure=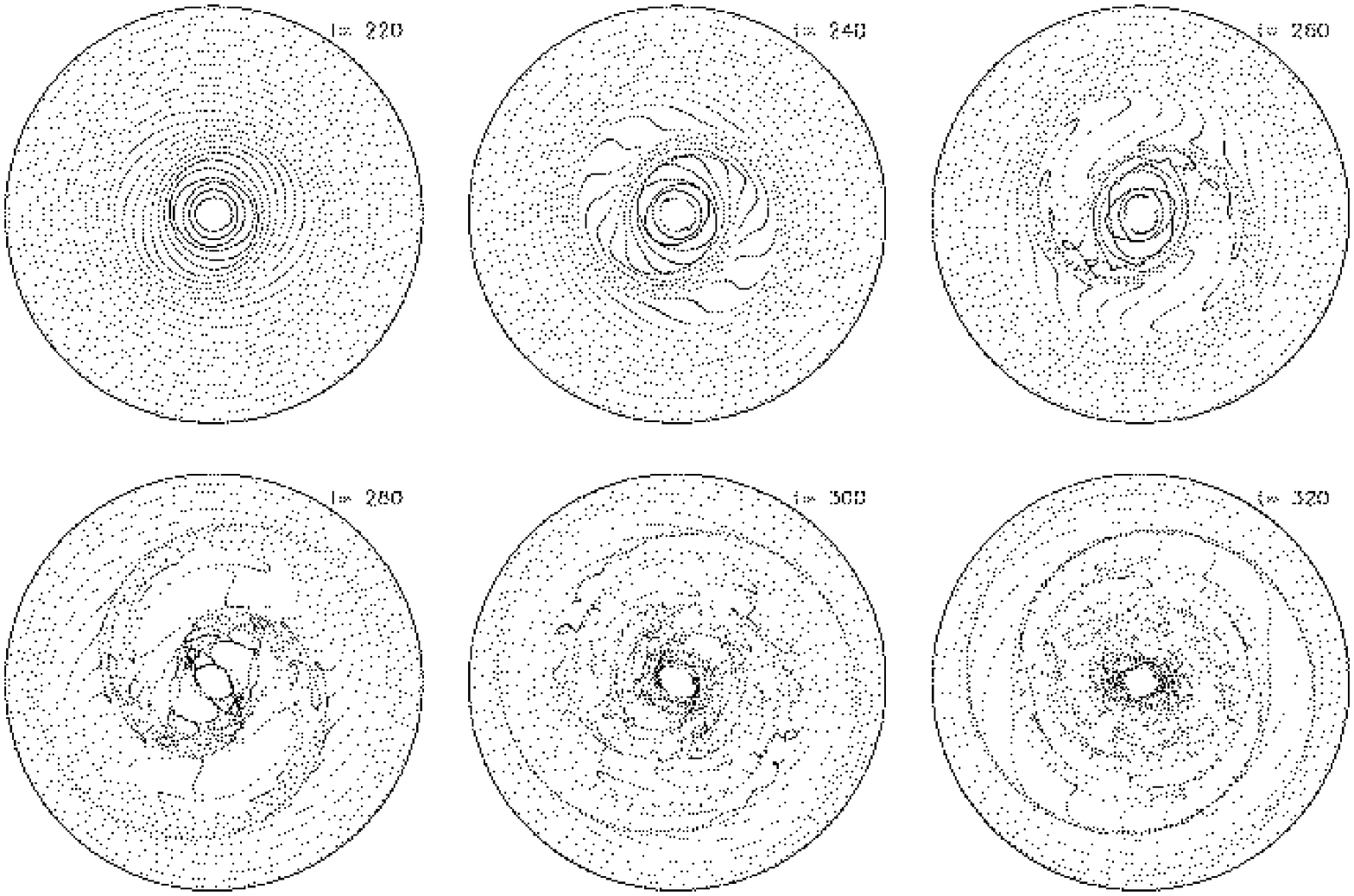,width=\hsize,angle=0,clip=}}
\caption{The evolution of rings of test particles on initially circular orbits in the simulation.  The rings can be thought of as showing the evolution of cold gas streams up until the moment they intersect.  Considerable mixing of the gas is clearly also to be expected.}
\end{figure}

\section{Radial mixing of stars and gas}

From these Figures, we conclude that (1) ``Lindblad resonance'' heating is 
minor and predominantly associated with small peculiar velocities, (2) wave 
damping begins well inside the resonances, and (3) that angular momentum 
changes at corotation dominate.

Evidently, the magnitude of the changes induced by a single spiral will depend strongly on its amplitude.  That shown in Figure~3 reaches a peak overdensity of $\sim 20\%$ -- 25\%, making it about twice the intensity of the strongest features seen in Figure 1.  Neither value seems unreasonably large for real galaxies, perhaps even rather conservative; if the typically larger intensity variations in K-band ({\it e.g.}\ Rix \& Zaritsky 1995; Block \& Puerari 1999) really scale as mass variations, then the angular momentum changes in real galaxies must exceed those reported here.

Assuming the time-dependence of this spiral to be typical of those in real galaxies, we find that a 
single spiral wave generally changes the angular momentum of a star near 
corotation by $\sim \pm20\%$.  Many stars are simply moved across corotation 
to another orbit with quite different angular momentum, but very similar 
radial action.  The changes near the Lindblad resonances, which cause most of 
the heating 
(Lynden-Bell \& Kalnajs 1972), are quite minor.

The churning of stars by multiple spiral events (Figure 2) has important 
consequences for the chemical abundance of stars within disks: gradients, 
which were established at the time stars formed must become shallower over 
time as a result of this activity.

It is important to realize that gas clouds are also scattered by the spiral 
perturbation.  To obtain some idea how the interstellar medium might respond to the spiral, we show in Figure~9 the evolution of rings of test particles, which began on initially circular orbits in the simulation shown in Figure~3.  The particles have no mass, and therefore do not contribute to the potential, but are integrated forward in the evolving potential from the disk of massive particles.  The rings illustrate how streams of cold gas would behave up until the time that rings either self-intersect, or cross another ring; from that moment, collisions (or pressure if the gas has a finite temperature) will intervene and alter the subsequent motion of gas.  Even so, the radial extent acquired by some (initially circular) rings indicates that substantial radial flows of gas are expected with in-going and out-going streams largely avoiding each other until about the time that the spiral saturates.

\section{Conclusions}

The main effect of a transient spiral wave is to churn the angular momentum 
distribution, with stars mostly changing places across corotation.  A 
succession of spiral waves of different pattern speeds causes a random walk 
in $L_z$ with a variable step size of rms $\sim 20\%$ of a star's initial $L_z$.  
It is 
therefore perfectly possible that the Sun 
has migrated by 2~kpc from its 
birthplace, as argued on other grounds by Wielen 
{\it et al.}\ (1996).

Remarkably, such substantial changes are not associated with any increase
 in 
random motion near corotation, which is perhaps why they have not previously 
attracted much attention.  They are also not really expected from previous work, 
such as the analyses presented by Lynden-Bell \& Kalnajs (1972) and Carlberg 
\& Sellwood (1985).  Those analyses assumed small departures from the unperturbed orbits, were also 
averaged over all phases, and focused on the limit of a steady wave, for which lasting changes are localized to resonances.  These assumptions lead to quite different expectations but, 
with hindsight, seem rather questionable.

The churning of the disk by spiral waves in this manner has extensive implications for chemical abundance gradients.  If stars can migrate considerably from their places of birth, abundance gradients established when they were formed will be diluted.  Furthermore, the gas in the disk will also be churned, and the coherent azimuthal dependence shown in Figure~9 shows that crossing of streams does not occur until considerable radial displacements have been achieved.

\acknowledgments We thank James Binney and Scott Tremaine for helpful 
discussions.  This work was supported by NSF grant AST-0098282 to JAS and by 
grant SFRH/BD/3444/2000 to MP from Funda\c{c}\~{a}o para a Ci\^{e}ncia e a 
Tecnologia, Portugal.

\end{document}